# MsDC-DEQ-Net: Deep Equilibrium Model (DEQ) with Multi-scale Dilated Convolution for Image Compressive Sensing (CS)

Youhao Yu,     Richard M. Dansereau[1]

Department of Systems and Computer Engineering, Carleton University, Ottawa, Canada

youhaoyu@cmail.carleton.ca        rdanse@sce.carleton.ca

**Abstract -** Compressive sensing (CS) is a technique that enables the recovery of sparse signals using fewer measurements than traditional sampling methods. To address the computational challenges of CS reconstruction, our objective is to develop an interpretable and concise neural network model for reconstructing natural images using CS. We achieve this by mapping one step of the iterative shrinkage thresholding algorithm (ISTA) to a deep network block, representing one iteration of ISTA. To enhance learning ability and incorporate structural diversity, we integrate aggregated residual transformations (ResNeXt) and squeeze-and-excitation (SE) mechanisms into the ISTA block. This block serves as a deep equilibrium layer, connected to a semi-tensor product network (STP-Net) for convenient sampling and providing an initial reconstruction. The resulting model, called MsDC-DEQ-Net, exhibits competitive performance compared to state-of-the-art network-based methods. It significantly reduces storage requirements compared to deep unrolling methods, using only one iteration block instead of multiple iterations. Unlike deep unrolling models, MsDC-DEQ-Net can be iteratively used, gradually improving reconstruction accuracy while considering computation trade-offs. Additionally, the model benefits from multi-scale dilated convolutions, further enhancing performance.

## 1. Introduction

Compressive sensing is a signal processing technique used to efficiently acquire signals that exhibit sparsity or compressibility in a sparse domain and this information can then be reconstructed back to the original domain with high probability [1], [2]. In compressive sensing, a signal is measured through a small number of linear projections obtained by multiplying the signal with a sensing matrix, typically a random matrix. For images, compressive sensing allows fewer measurements to be acquired than the number of pixels in an image, making it particularly advantageous as it can significantly reduce storage requirements. Reconstruction of the image is usually obtained using algorithms that leverage the sparsity of the signal, such as sparse optimization or convex optimization techniques, but these techniques are often computationally expensive and slow to converge. Compressive sensing finds applications in various fields, including magnetic resonance imaging (MRI) [3], radar signal sampling [4], cryptosystems [5], snapshot imaging [6] and video sensing [7], [8]. It proves especially useful when dealing with large amounts of data, as it can lead to significant reductions in storage and processing requirements.

A multitude of optimization-based CS reconstruction methods have been developed. One such method is Basis Pursuit (BP), an algorithm that tackles the underdetermined linear system by finding the sparsest solution [9]. It assumes signal sparsity in a specific basis and solves a convex optimization problem to determine the sparsest representation. Another iterative algorithm, Iterative Hard Thresholding (IHT), updates the signal estimate by applying thresholding at each iteration [10]. The algorithm computes a gradient descent step and enforces sparsity through thresholding, where only the $k$ largest coefficients ($k$ being the desired sparsity level) are retained, while the rest are set to zero. Compressive Sampling Matching Pursuit (CoSaMP), another iterative algorithm, iteratively refines the signal estimate by selecting the support that best aligns with the measurements [11]. At each iteration, CoSaMP identifies the $k$ largest entries in the product of the sensing matrix $\boldsymbol{\Phi}$ and the current residual. It then solves a least-squares problem to obtain the coefficients of the selected atoms. Approximate Message Passing (AMP), an iterative algorithm utilizing a message-passing framework, estimates the signal by combining the current estimate with the noisy measurements and applying a soft thresholding operator [12]. The result undergoes linear combination and soft thresholding at each iteration. An optimization algorithm frequently employed in sparse signal recovery and regression problems is Iterative Shrinkage Thresholding Algorithm (ISTA). It is a variant of the proximal gradient descent algorithm [13]. ISTA iteratively updates the estimate of the sparse signal by taking a gradient step and subsequently applying thresholding to promote sparsity. The thresholding operation sets small entries to zero, and the degree of sparsity is controlled by the threshold value. These methods, while effective, suffer from high computational complexity due to the necessity of multiple iterations to achieve convergence. Additionally, some parameters require careful tuning for optimal performance.

---

[1] Corresponding author: Richard M. Dansereau

In recent years, neural network-based CS reconstruction methods have gained popularity. These methods leverage the ability of neural networks to learn complex, non-linear mappings between compressed measurements and reconstructed signals. Unlike traditional methods mentioned earlier, these non-iterative network-based approaches significantly reduce computational requirements while achieving impressive reconstruction performance [14]. Most network-based methods are trained as black boxes, harnessing the powerful learning capacity of deep networks but lacking insights from a CS perspective. On the other hand, optimization methods involve iterating over parameters to minimize a loss function. Deep unrolling methods (DUM) in machine learning can be seen as incorporating insights from iterations in optimization methods [14]. In deep unrolling, a fixed number of architecturally identical blocks are utilized, where the output of each block serves as input to the next. This can be interpreted as a form of iteration, where the block is applied iteratively (typically 5 to 10 times) to capture longer-term dependencies [15]. While more iteration blocks often yield improved performance, training such models consumes substantial memory, potentially leading to out-of-memory issues [16]. Deep equilibrium models (DEQ) belong to a class of deep learning models that employ fixed-point iteration schemes to learn stable equilibrium points, corresponding to optimal solutions for given optimization problems [17]. DEQ has been applied to compressive sensing reconstruction by formulating the problem as an optimization task solvable through fixed-point iteration schemes [15], [16], [18]. By utilizing a neural network to learn the fixed-point iteration scheme, DEQ demonstrates the ability to reconstruct images accurately and efficiently from compressive measurements.

This paper presents the design of a deep equilibrium model, named MsDC-DEQ-Net, for image compressive sensing, incorporating multi-scale dilated convolutions. The model consists of two key components. Firstly, we utilize semi-tensor product theory (STP) to enable direct compressed measurement and initial reconstruction without the need for image block processing. This approach avoids block artifacts in reconstruction and employs a learnable measurement matrix to capture essential signal information. Secondly, we construct a deep equilibrium layer based on one iteration of the ISTA algorithm, mapping $l_1$ norm optimization for CS reconstruction into a deep network. We incorporate aggregated residual transformations (ResNeXt) [19] to enhance performance and employ the squeeze-and-excitation network (SENet) [20] to remove redundancy and enhance valuable information. Compared to deep unrolling methods, our proposed MsDC-DEQ-Net possesses fewer learnable parameters and offers a trade-off between accuracy and computation.

The main contributions of this work can be summarized as follows. (1) By leveraging semi-tensor product (STP) without image block partitioning, we achieve image sampling and initial reconstruction, mitigating block artifacts and using a learnable measurement matrix that captures critical signal information. (2) We map one iteration of the ISTA algorithm into a network layer, referred to as the ISTA block, and enhance its performance by incorporating the ResNeXt and SENet structures. Additionally, we apply multi-scale dilated convolutional layers to further improve performance. (3) To address the issue of large model size associated with deep unrolling methods, we employ the ISTA block to construct a deep equilibrium model. This allows for multiple applications of the trained model, enabling multiple ISTA iterations and a continuous improvement in reconstruction performance.

## 2. Related Work

Network-based compressive sensing image reconstruction methods can be broadly classified into two categories: deep non-unfolding networks and deep unfolding networks [21]. Each category encompasses specific methods that vary in terms of network architectures, loss functions, regularization techniques, and other details. The following section provides a concise overview of both reconstruction methods. Subsequently, we delve into the exploration of the deep equilibrium model, focusing on its relevance to our own research.

**A. Network-based compressive sensing image reconstruction methods**

**1) Deep non-unfolding networks:** Deep non-unfolding networks are deep neural networks trained to directly reconstruct under-sampled images from compressed measurements, without explicitly modeling the image acquisition process. These methods typically involve training a deep neural network using large-scale datasets of natural images to learn a mapping from under-sampled measurements to the fully sampled image domain.

Usually, these reconstructed images often lack fine details, particularly at low measurement rates. To address this issue, the dual-path attention network (DPA-Net) [22] employs two paths. The structure path focuses on reconstructing the dominant structural components, while the texture path recovers the remaining texture details. An attention module is utilized to transmit structure information to the texture path. To reduce the number of parameters, block-based compressed sensing (BCS) is often employed for sampling and reconstructing small image blocks. The sampling and initial reconstruction are achieved using a sampling and whole image denoising network based on a generative adversarial network (SWDGAN) [23]. Non-overlapping blocks are segmented from the original images, and a fully connected layer is utilized for sampling and initial reconstruction. A whole image dense residual denoising module is then applied to further improve the reconstruction quality. The generator and discriminator are trained alternately to obtain an optimal model. Similar to DPA-Net, the parallel enhanced network (PE-Net) [24] consists of two reconstruction networks. The basic network produces the initial reconstruction, while the enhanced network progressively refines details by utilizing information from sub-modules of the basic network. The final reconstruction is the cumulative result of the two parallel networks.

These models all employ block-based compressed sensing (BCS) methods, where large-scale images are processed in a block-by-block manner. In contrast, the semi-tensor product network (STP-Net) [25] treats the image as a whole without segmentation. Leveraging semi-tensor product theory, an image can be directly sampled using a small-sized measurement matrix through matrix multiplication. The initial reconstruction can also be obtained in a similar manner. It is worth noting that these models are trained as black boxes, lacking insights from the compressive sensing domain.

**2) Deep unfolding networks:** Deep unfolding networks aim to unfold the iterative optimization process of traditional compressive sensing (CS) reconstruction algorithms, such as ISTA, into a single end-to-end trainable deep neural network. These networks explicitly model the physics of the image acquisition process and learn a mapping from under-sampled measurements to the original image by iteratively updating the reconstructed image estimate.

The ISTA-Net [14] approach solves CS reconstruction using the ISTA algorithm by casting it into a deep network form. This allows it to benefit from the structural insights of traditional optimization-based methods while maintaining the fast solution speed of neural networks. Nonlinear convolutional layers are employed to solve the proximal mapping associated with the sparsity-inducing regularizer. ISTA-Net's network design is well-defined, providing interpretability and allowing for structural diversity originating from the compressive sensing domain. In contrast to ISTA-Net, the optimization-inspired explicable deep network (OPINE-Net) [26] utilizes a data-driven adaptively learned matrix instead of generating the sampling matrix with a fixed random Gaussian matrix. Although the sensing matrix can take other forms, such as *noiselets* coefficients without additional memory storage or multiplications as shown in [27]–[29], data-driven methods are expected to have better performance for learnable classes of data. OPINE-Net adopts the framework of ISTA-Net, with each of its 9 blocks corresponding to one iteration in the traditional ISTA algorithm. Notably, all blocks share the same weights without affecting the final reconstruction performance. AMP-Net [30] unfolds the iterative denoising process of the approximate message passing (AMP) algorithm. Similar to OPINE-Net, AMP-Net consists of 9 AMP denoising iteration blocks, and the sampling matrix is trainable. Additionally, AMP-Net integrates deblocking modules to eliminate blocking artifacts. STP-ISTA-Net, introduced in [16], combines STP-Net and ISTA-Net by connecting the output of the former as a better initial reconstruction for the latter. This model uses five iteration blocks, fewer than the aforementioned models, while still achieving competitive performance.

**B. Deep equilibrium model**

In [31], the authors observe that the hidden layers of many existing sequence models converge to a fixed point. To address this, they propose the deep equilibrium model (DEQ), which allows for finding the equilibrium point directly via root-finding. This method is equivalent to running an infinite-depth (weight-tied) network, and the equilibrium point can be backpropagated using implicit differentiation. The DEQ model requires only constant memory.

While deep unrolling methods like ISTA-Net, OPINE-Net, and AMP-Net achieve good performance by simulating a fixed number of iterations of an optimization method in their architectures, the number of iterations must be limited due to the difficulty of training large-sized networks. Additionally, significant errors arise when expecting more optimization iterations through multiple applications of the trained model [15]. In contrast, the DEQ model can be executed for more optimization iterations, leading to consistent improvements in reconstruction quality while requiring only constant memory in both training and testing [15]. There exists a trade-off between reconstruction quality and computation.

The DEQ model has been applied to inverse problems in imaging [15], [16] and video snapshot compressive imaging (SCI) reconstruction [18]. Although the DEQ model simplifies the structure compared to deep unrolling methods, a bottleneck arises with single-scale convolutions, limiting the ability to extract and propagate useful information. Dilated convolutions are widely used in various domains such as image denoising [32], feature detection [33], image super-resolution [34] and CS reconstruction [35]. Dilated convolutions expand the receptive field without increasing the number of parameters, thereby maintaining the same amount of computation. This inspires us to extract features of different scales using a model that incorporates multiple convolution channels in parallel, with each channel having different dilation factors.

In comparison to deep non-unfolding networks, the architecture of MsDC-DEQ-Net offers good interpretability as it borrows insights from traditional optimization methods. MsDC-DEQ-Net also allows for structural diversity in its model design, providing ample room for optimizing network structures. Compared to deep unfolding networks, MsDC-DEQ-Net only requires one iteration block, significantly reducing memory requirements and addressing computation issues in large-scale models [15]. Extensive experiments demonstrate that MsDC-DEQ-Net achieves competitive performance compared to existing network-based CS image reconstruction methods.

## 3. Proposed MsDC-DEQ-Net for Image Compressive Sensing

In this section, we will first introduce the relevant concepts that we have utilized and then provide a detailed explanation of the design of the proposed MsDC-DEQ-Net.

### 3.1 ISTA Optimization for CS

Suppose the original signal $x \in \mathbb{R}^N$ is CS measured by a linear random projection $\boldsymbol{\Phi} \in \mathbb{R}^{M \times N}$ giving measurements $y \in \mathbb{R}^M$ as

$$y = \boldsymbol{\Phi} x. \tag{1}$$

were $M \ll N$ and the compressive sensing ratio is given as $M/N$. The purpose of CS reconstruction is to infer $x$ from $y$. Traditionally, we obtain the reconstruction by solving the optimization problem

$$\min_{x} \left\{ \frac{1}{2} \|\boldsymbol{\Phi} x - y\|_2^2 + \lambda \|\boldsymbol{\Psi} x\|_1 \right\}, \tag{2}$$

where $\boldsymbol{\Psi}$ is some sparse transform and the result of $\boldsymbol{\Psi} x$ is the coefficients of $x$ in a sparse domain. The sparsity of $\boldsymbol{\Psi} x$ is encouraged by the $l_1$ norm with regularization parameter $\lambda$.

The problem in (1) can be solved using various optimization algorithms, such as the iterative shrinkage thresholding algorithm (ISTA) [13], alternative direction method of multipliers (ADMM) [36], and approximate message passing (AMP) [12]. In this paper, we adopt the ISTA algorithm for simplicity. ISTA is a widely used first-order proximal method and is particularly suitable for solving linear inverse problems [14]. Each iteration of the ISTA algorithm consists of the two steps

$$r^{(k)} = x^{(k-1)} - \rho \boldsymbol{\Phi}^T (\boldsymbol{\Phi} x^{(k-1)} - y), \tag{3}$$

$$x^{(k)} = \boldsymbol{\Psi}^{-1} \left\{ \left( |\boldsymbol{\Psi} r^{(k)}| - \lambda \right)_+ sgn(\boldsymbol{\Psi} r^{(k)}) \right\}, \tag{4}$$

where $r$ is the immediate reconstruction and $\rho$ is the step size [13], [14]. The subscript + in (4) takes the positive part, setting any negative part to zero. In (3) is the gradient descent step where the estimate of the reconstructed signal $x$ is updated by taking a step in the direction of the negative gradient of the objective function using the data fidelity term $\frac{1}{2}\|\boldsymbol{\Phi} x - y\|_2^2$ from (2). This step is essentially a gradient descent update, where the step size is chosen based on the Lipschitz constant of the gradient of the objective function [13]. In (4) is the soft-threshold step where the soft-thresholding operator is applied to the current estimate of the signal in its sparse domain. To be specific, the operator first sets the elements whose absolute values are below a certain threshold to zero, and then shrinks the nonzero coefficients toward zero by the threshold. These two steps are repeated iteratively until convergence is achieved or a maximum number of iterations is reached. The ISTA algorithm is commonly used for solving sparse linear regression and compressed sensing problems [13].

### 3.2 Semi-tensor Product (STP)

According to STP theory, a small matrix $\boldsymbol{\Phi}$ can be multiplied with a tall vector $x$ as follows

$$y = \boldsymbol{\Phi}(t) \ltimes x, \tag{5}$$

where $\boldsymbol{\Phi}(t) \in \mathbb{R}^{\frac{M}{t} \times \frac{N}{t}}$, $\ltimes$ is the left product operator for STP, and $t$ is a shrinkage factor chosen as a common divisor of $M$ and $N$ [37]–[39]. The operation in (5) is equivalent to

$$y = \{\boldsymbol{\Phi}(t) \otimes I_t\} x, \tag{6}$$

where $\otimes$ is the Kronecker product of matrices and $I_t \in \mathbb{R}^{t \times t}$ is an identity matrix. When $t = \sqrt{N}$, (6) can be written in matrix form [25] as

$$Y = X \cdot \boldsymbol{\Phi}(t)^T, \tag{7}$$

where $X \in \mathbb{R}^{\sqrt{N} \times \sqrt{N}}$, $Y \in \mathbb{R}^{\sqrt{N} \times \frac{M}{\sqrt{N}}}$ and $\boldsymbol{\Phi}(t) \in \mathbb{R}^{\frac{M}{\sqrt{N}} \times \sqrt{N}}$. Here, we reshape the vectors $x$ and $y$ to be matrices maintaining column-wise order. The square matrix $X$ is easy to associate with a square image.

If $\boldsymbol{\Phi}(t)$ satisfies the restricted isometry property (RIP) [40], it can be used as a measurement matrix since mutual coherence of $\boldsymbol{\Phi}(t) \ltimes \boldsymbol{\Psi}$ still satisfies the RIP [41]. The application of STP brings significant convenience to compressive sensing. In [41], the authors perform column-wise measurements of an image, reducing the dimensions of the measurement matrix to $1/t^2$ compared to conventional CS. This reduction greatly reduces the memory footprint.

If we consider an image with a size of 256×256 and directly apply compressive sensing measurements at a ratio of 10%, the size of the measurement matrix $\boldsymbol{\Phi}(1)$ would be 6554×65536. However, to conserve memory, block-based compressed sensing is commonly employed. In approaches such as [14], [26], [30], [42]–[44], the image is divided into smaller blocks, typically with a size of 33×33. These blocks are then vectorized and measured individually. In the case of a 10% CS ratio, this means a measurement matrix with a size of 109×1089 would be required, but at the disadvantage of creating blocking artifacts when these blocks are reassembled. As shown in [27], block-based sensing is less efficient compared with global sensing in recovery performance because a measurement in the block-based approach has information only about the block, while a measurement in the global approach has information about the whole image.

Using the same 256×256 image, if the image is vectorized directly without breaking it into smaller blocks and STP is applied, when $t$ is chosen as 256 the measurement matrix $\boldsymbol{\Phi}(256)$ only has size 26×256. So, the measurement matrix size is smaller than the block-based method that used blocks of size 33×33 instead of the full image of size 256×256.

There can be fluctuations in the resulting mutual coherence of different generated measurement matrices, especially when the value of $t$ is large, resulting in a decreased probability of satisfying the restricted isometry property (RIP) [41]. This compels us to adopt a larger-sized measurement matrix from a smaller value of $t$. In accordance with [25], an image can be measured using two steps with two larger measurement matrices as

$$\boldsymbol{Y} = \boldsymbol{\Phi}_1(t) \cdot \boldsymbol{X} \cdot \boldsymbol{\Phi}_2(t)^T, \tag{8}$$

where the combined steps results in fewer resulting measurements in $\boldsymbol{Y}$. For instance, to achieve a CS ratio equal of 10%, the size of the two measurement matrices $\boldsymbol{\Phi}_1(256)$ and $\boldsymbol{\Phi}_2(256)$ are set to 81×256, since $(81/256)^2 \approx 10\%$. The memory footprint they occupy is still smaller than a 33×33 block-based method. Here, $\boldsymbol{\Phi}_1(t)$ and $\boldsymbol{\Phi}_2(t)$ can even be selected to be the same. Based on (8), we build STP-Net as shown in Figure 1 [25], which can measure an image directly without segmenting into blocks and provides an initial reconstruction. Mea1 and Mea2 correspond to $\boldsymbol{\Phi}_1(t)$ and $\boldsymbol{\Phi}_2(t)$, showing the two measurement steps. Rec1 and Rec2 are the inverse operation of Mea1 and Mea2 for a total of four matrices for the measurement and initial reconstruction phases.

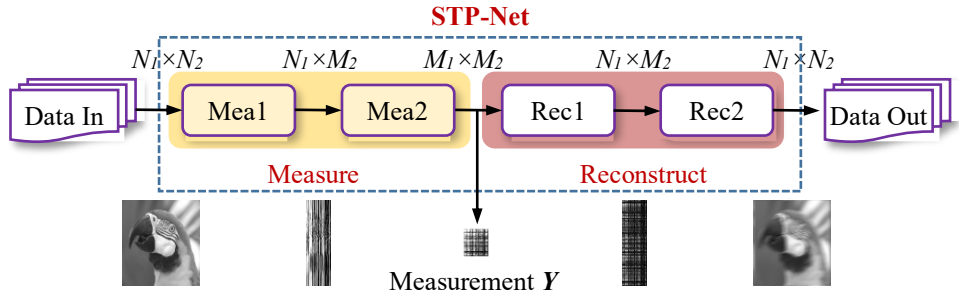

Figure 1: STP-Net [25]

### 3.3 Structure of ISTA Iteration Block

To implement (3), we build the immediate reconstruction block (IRB) as shown in Figure 2 by means of STP-Net.

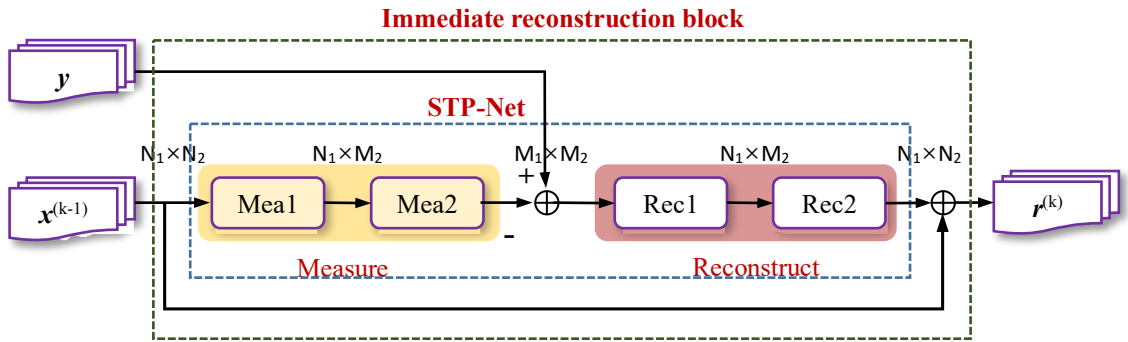

Figure 2: Immediate reconstruction block (IRB)

The residuals of an image are known to exhibit higher compressibility [14], and incorporating residual learning can facilitate the training of deeper networks [45]. Let's assume that $x^{(k)}$ comprises three components: the immediate reconstruction result $r^{(k)}$, the high-frequency components $h^{(k)}$ missing in $r^{(k)}$ found using an appropriate operator, and any remaining noise $n^{(k)}$ embedded in $x^{(k)}$ [14]. We can express $x^{(k)}$ as

$$x^{(k)} = r^{(k)} + h^{(k)} + n^{(k)}. \tag{9}$$

Then, (4) can be reformulated as

$$x^{(k)} = r^{(k)} + \mathcal{H}\left\{\Psi^{-1}\left(\left(\left|\Psi\left(\mathcal{D}(r^{(k)})\right)\right| - \lambda\right)_{+} sgn\left(\Psi\left(\mathcal{D}(r^{(k)})\right)\right)\right)\right\} + n^{(k)}, \tag{10}$$

where $\mathcal{D}$ represents a denoising operation and $\mathcal{H}$ represents a high-pass filter [14].

Since network-based methods allows for structural diversity originating from the CS domain [14], we construct the ISTA block, as shown in Figure 3, which implements (3) and (10). All "Conv" layers represent convolutional layers with a filter size of 3×3. The top line in Figure 3 represents the output feature sizes, while the second line represents the number of features. The blocks $\mathcal{F}^k$ and $\tilde{\mathcal{F}}^k$ correspond to $\Psi$ and $\Psi^{-1}$, respectively, implementing the sparse transform and its inverse operation [14]. The second output of Figure 3 is enforced to be zero, indicating that a signal passes through $\mathcal{F}^k$ and $\tilde{\mathcal{F}}^k$ without any changes.

MsDC (multi-scale dilated convolution) is illustrated in Figure 4, which consists of multiple branches with different dilation factors ranging from 1 to 7. This enables the extraction of both structural and detailed information from natural images, as different dilation factors capture different levels of information [32].

The denoise layer $\mathcal{D}$ and high-pass filter layer $\mathcal{H}$, depicted in Figure 5(a) and (b) respectively, are inspired by the aggregated residual transformations for deep neural networks (ResNeXt) [19] and the squeeze-and-excitation (SE) network [20]. ResNeXt increases the number of branches in a residual block, while the SE network adaptively recalibrates channel-wise feature responses. Since MsDC extracts multi-features from $r^{(k)}$, the structure of the SE network is expected to enhance important features, while the structure of ResNeXt is expected to improve the performance of the denoise layer and high-pass filter.

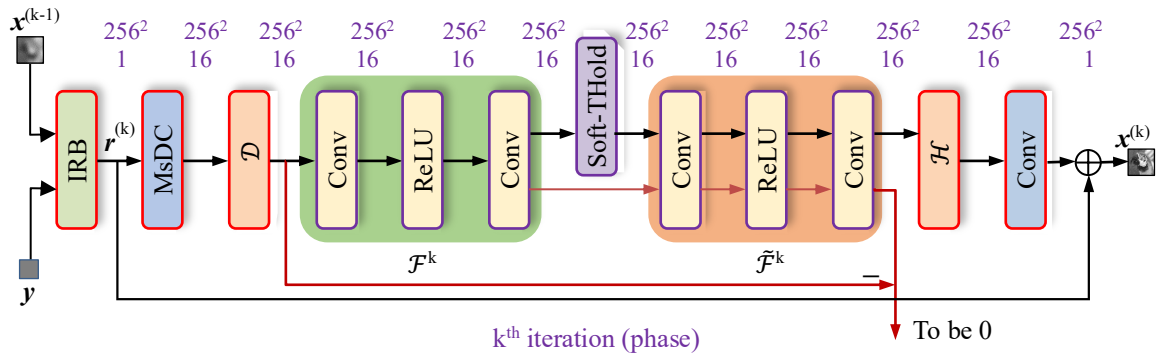

Figure 3: ISTA block

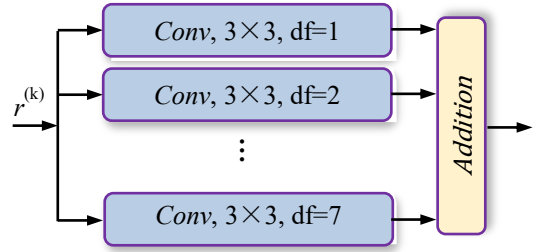

Figure 4: Multi-scale dilated convolution layer (df: dilation factor)

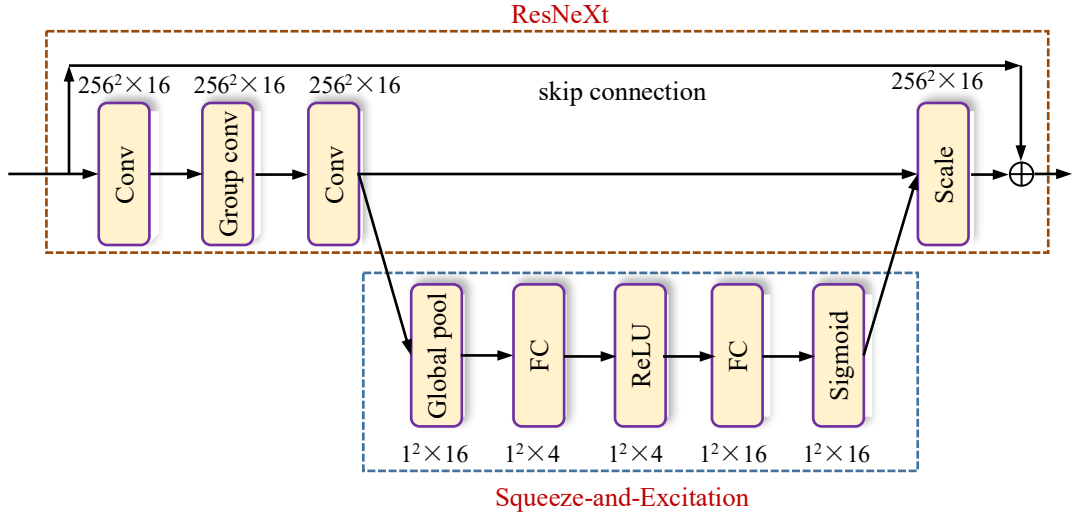

(a) Denoise layer $\mathcal{D}$ [19], [20]

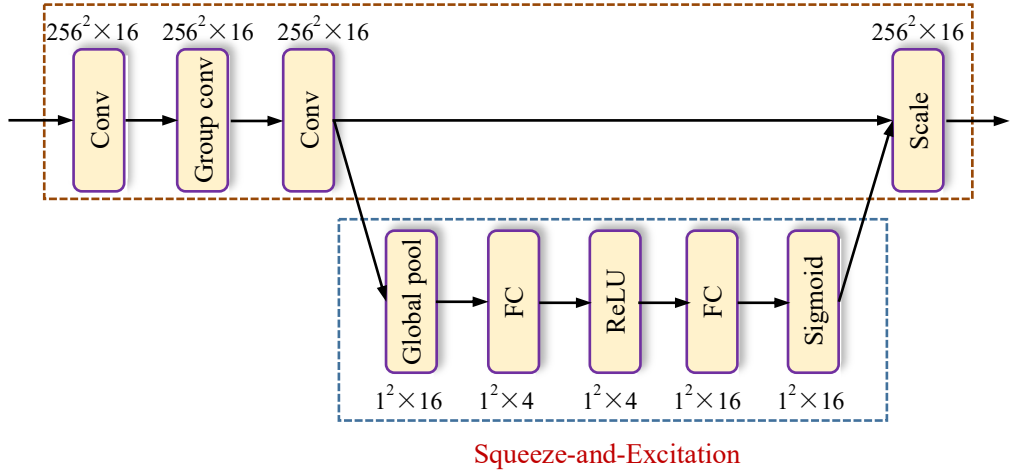

(b) High pass layer $\mathcal{H}$ [19], [20]

Figure 5: Structure of denoise layer and high pass layer

### 3.4 Proposed MsDC-DEQ-Net

From Figure 3, an ISTA iteration can be described as

$$x^{(k)} = f(x^{(k-1)}, y), \qquad (11)$$

where $f(\cdot)$ represents the operations of all the layers in Figure 3. Equation (11) merges (3) and (10) into one equation.

According to the ISTA algorithm, after a certain number of iterations $x^{(k-1)}$ will approach $x^{(k)}$, converging to an equilibrium point. This observation aligns with the concept of an equilibrium model, which has a fixed point $x^*$ represented as

$$x^* = f(x^*, y), \qquad (12)$$

where $y$ serves as the input injection, playing a crucial role in ensuring that the equilibrium point aligns with the original signal [17]. Therefore, we can consider the ISTA block as an equilibrium layer within the model.

The proposed MsDC-DEQ-Net for image compressive sensing is illustrated in Figure 6. Both the STP-Net and the equilibrium layer are jointly trained. The STP-Net provides the measurement $y$ as the input injection for the equilibrium layer, along with the initial reconstruction $z_0$, which serves as a suitable starting point for the iterative solution within the equilibrium layer. Figure 6 shows three outputs. (1) Output1 aims to approach the original images. (2) Output2 ensures the reversibility of the sparse transform, meaning that the signal remains unchanged as it passes through $\mathcal{F}^k$ and $\tilde{\mathcal{F}}^k$. Therefore, Output2 should equal zero, as depicted in Figure 3. (3) Output3 focuses on the initial reconstruction, striving to approach the original image, as it significantly contributes to solving the equilibrium point efficiently. The parameters of the MsDC-DEQ-Net are optimized by minimizing the half mean square error (HMSE) between the outputs and the expected signals.

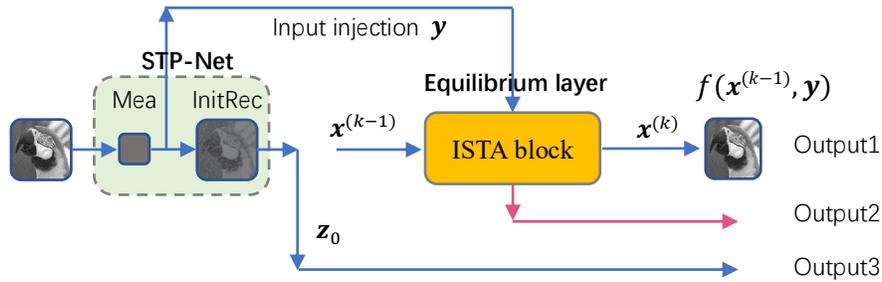

Figure 6: MsDC-DEQ-Net

## 4 Experimental Results

The ILSVRC2014 ImageNet dataset consists of 1.2 million images with 1,000 object categories and is commonly used in computer vision competitions [46]. For our experiments, we randomly selected 20,000 natural images from this dataset, with 14,000 images used for training, 3,000 for validation, and 3,000 for testing. Each image was cropped to a central 256×256 region and converted to 8-bit grayscale. During training, we employed the Adam solver with a learning rate of 1e-5 and a minibatch size of 16. To evaluate the performance of our model, we used two widely-used benchmark datasets: Set11 [42] and BSD68 [47]. Set11 contains 11 grayscale images, while BSD68 contains 68 grayscale images. The reconstruction results were reported for compression ratios of 1%, 4%, 10%, and 25%. The peak signal-to-noise ratio (PSNR) and structural similarity index (SSIM) were used as evaluation criteria.

All experiments were conducted on MATLAB R2019a, running on a computer with an Intel i7-8700K CPU operating at 3.7 GHz, a GeForce GTX 1080 GPU, and 16 GB RAM. To find the fixed point of the DEQ model, we employed the Anderson acceleration method to improve convergence and prevent divergence. This method determines the promising direction for iterations by updating the input of the deep equilibrium layer with a linear combination of previous outputs. During test time, we set the number of iterations to 50 for the Anderson acceleration method.

### 4.1 Performance Comparison

We conducted a comparison between our proposed MsDC-DEQ-Net and nine recent state-of-the-art image CS methods, namely STP-Net [25], DPA-Net [22], SWDGAN [23], PE-Net [24], ISTA-Net$^+$ [26], OPINE-Net [26], AMP-Net [30], STP-ISTA-Net [16] and STP-DEQ-Net [16]. The first four are deep non-unfolding networks, while the next four are deep unfolding networks. We evaluated the average PSNR and SSIM reconstruction performance on the Set11 dataset across four compressive sensing ratios, as summarized in Table 1. The results for the other methods were obtained from their respective papers.

We anticipate that the proposed technique will perform well compared to the listed competing techniques. For the proposed technique, Fig. 5(a) and (b) show that we have incorporated aspects of ResNeXt [19] and the Squeeze-and-Excitation network [20] into our model. These two networks have good performance, so incorporating these into our proposed network is expected to perform well compared with techniques like ISTA-NET[+]. From Table 1, it is evident that our proposed model outperforms the other methods with higher PSNR and SSIM scores, particularly at the extremely low CS ratio of 1%. Even at CS ratios of 4% and 10%, our model still achieves superior PSNR. Although our proposed model has slightly lower performance at a CS ratio of 25%, it remains competitive. By increasing the number of iterations in the Anderson acceleration method, we can obtain even better results.

To assess the generalizability of our model, we also compared it with other methods on the larger BSD68 dataset. As shown in Table 2, our proposed model achieves the best performance at CS ratios of 10% and 25%. It achieves the second-best performance at a CS ratio of 4%. At the extremely low CS ratio of 1%, our proposed model exhibits higher PSNR but slightly inferior SSIM compared to SWDGAN and AMP-Net.

**Table 1:** Mean PSNR (dB) and SSIM comparisons on Set11 at different CS ratio [16]

| Algorithm | CR=1% | | CR=4% | | CR=10% | | CR=25% | |
|---|---|---|---|---|---|---|---|---|
| | PSNR | SSIM | PSNR | SSIM | PSNR | SSIM | PSNR | SSIM |
| **STP-Net** [25] | 20.73 | 0.5111 | 23.39 | 0.6310 | 26.08 | 0.7679 | 30.06 | 0.8843 |
| **DPA-Net** [22] | 18.05 | 0.5011 | 23.50 | 0.7205 | 26.99 | 0.8354 | 31.74 | 0.9238 |
| **SWDGAN** [23] | 21.01 | 0.5410 | 24.67 | 0.6410 | 28.45 | 0.8570 | 33.45 | 0.9300 |
| **PE-Net** [24] | 20.74 | 0.4909 | 25.19 | **0.7586** | 28.58 | 0.8701 | 33.23 | 0.9407 |
| **ISTA-Net[+]** [26] | 17.42 | 0.4029 | 21.32 | 0.6037 | 26.64 | 0.8087 | 32.59 | 0.9254 |
| **OPINE-Net** [26] | 19.87 | 0.5070 | 25.04 | 0.7730 | 29.33 | **0.8825** | 34.44 | **0.9491** |
| **AMP-Net** [30] | 20.20 | 0.5581 | 25.26 | 0.7722 | 29.40 | 0.8779 | **34.63** | 0.9481 |
| **STP-ISTA-Net** [16] | 21.32 | 0.5529 | 25.47 | 0.7152 | 29.01 | 0.8438 | 33.72 | 0.9327 |
| **STP-DEQ-Net** [16] | 21.24 | 0.5565 | 24.67 | 0.7028 | 28.96 | 0.8495 | 32.63 | 0.9331 |
| **MsDC -DEQ-Net** | **21.43** | **0.5642** | **25.55** | 0.7303 | **29.96** | 0.8723 | 34.09 | 0.9446 |

(The best performance is labeled in bold.)

**Table 2:** Mean PSNR (dB) and SSIM comparisons on BSD68 at different CS ratio

| Algorithm | CR=1% | | CR=4% | | CR=10% | | CR=25% | |
|---|---|---|---|---|---|---|---|---|
| | PSNR | SSIM | PSNR | SSIM | PSNR | SSIM | PSNR | SSIM |
| **STP-Net** [25] | 21.06 | 0.4639 | 23.11 | 0.5804 | 25.11 | 0.7091 | 28.48 | 0.8473 |
| **DPA-Net** [22] | 18.98 | 0.4643 | 23.27 | 0.6096 | 25.57 | 0.7267 | 29.68 | 0.8763 |
| **SWDGAN** [23] | 22.28 | **0.5600** | 24.88 | 0.6700 | 27.29 | 0.7890 | 30.96 | 0.9100 |
| **PENet** [24] | 22.10 | 0.4820 | 25.06 | 0.6711 | 27.38 | 0.7960 | 30.97 | 0.9053 |
| **ISTA-Net[+]** [26] | 19.14 | 0.4158 | 22.17 | 0.5486 | 25.32 | 0.7022 | 29.36 | 0.8525 |
| **OPINE-Net** [26] | 21.80 | 0.4972 | 24.87 | 0.6709 | 27.54 | 0.7966 | 31.28 | 0.9034 |
| **AMP-Net** [30] | 22.28 | 0.5387 | **25.26** | **0.6760** | 27.86 | 0.7926 | 31.74 | 0.9048 |
| **STP-ISTA-Net** [16] | 21.87 | 0.5046 | 25.12 | 0.6639 | 27.94 | 0.7932 | 31.35 | 0.8956 |
| **STP-DEQ-Net** [16] | 21.38 | 0.4882 | 23.74 | 0.6209 | 26.58 | 0.7615 | 30.48 | 0.8966 |
| **MsDC-DEQ-Net** | **22.53** | 0.5316 | 25.18 | 0.6724 | **28.24** | **0.7988** | **32.17** | **0.9136** |

(The best performance is labeled in bold.)

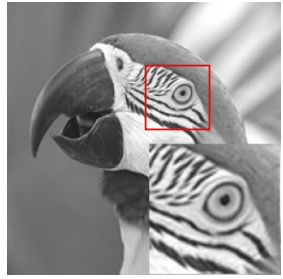

Figure 7: Original image of parrot with zoomed in region around the eye.

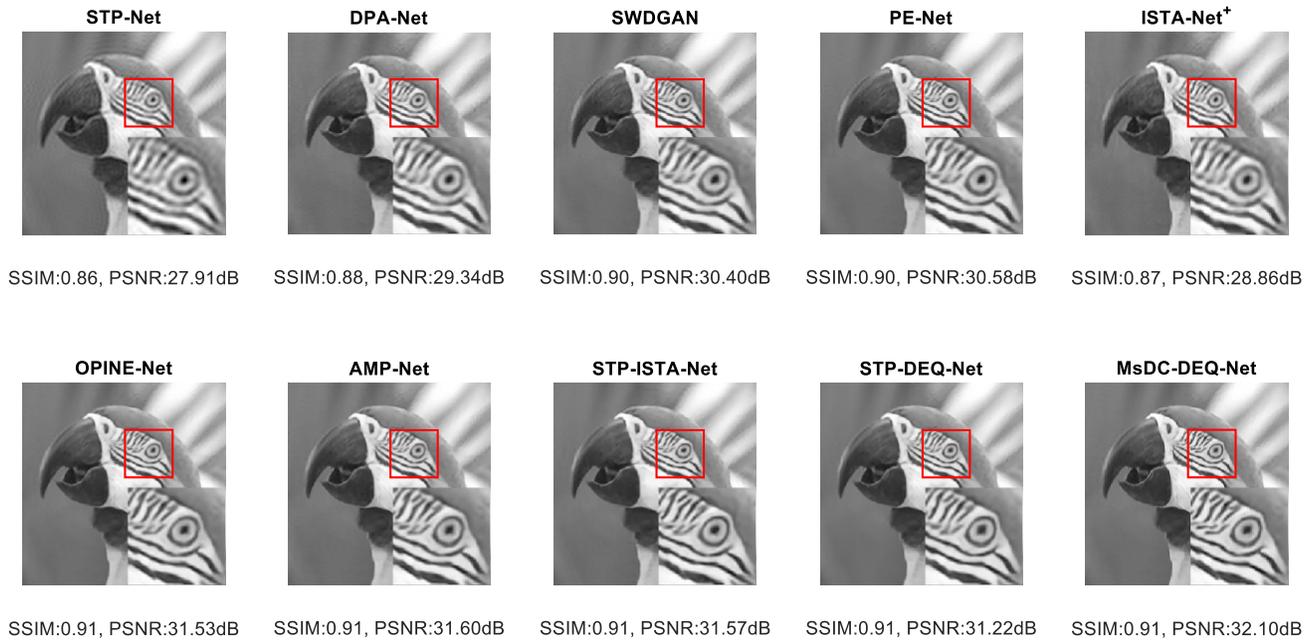

Figure 8: Reconstruction of parrot image with 10 different methods (Corresponds with Table 1 when CR=10%)

For visualization purposes, Figure 7 shows an original image of a parrot Figure 8 shows the CS reconstruction with different techniques. By zooming in on local area around the parrot's eye, it is seen that the proposed method has more realistic reconstruction than the other methods shown. For instance, the stripes around the eye appear to be better reconstructed in the proposed approach, which we expect is due to the multi-scale dilated convolution network introduced into the model.

**4.2 Ablation Studies**

The multi-scale dilated convolution model has demonstrated remarkable performance due to its ability to extract features at different scales from an image, where the combination of these features contributes to better reconstruction [35]. Previous studies [32]–[35] have often utilized dilation factors of 1, 2, and 3. In our proposed model, we intentionally incorporated seven branches with different dilation factors, as depicted in Figure 4, to better observe the impact of these factors. The branches are labeled from 1 to 7 based on their corresponding dilation factor values. To facilitate training, at the outset, we assigned the seven branches with the same parameter values as a pre-trained STP-DEQ-Net, with the exception of the dilation factor. We conducted several ablation studies to analyze the effects, as shown in Tables 3 through 7 following [48].

The results in Table 3 indicate that when all seven parallel branches are connected (denoted as "under all"), the model effectively learns the residual and achieves good performance. Conversely, when no branches are connected (denoted as "under none"), the model essentially produces an immediate reconstruction $r^{(k)}$, which requires improvement in terms of quality. Notably, when only one branch is connected, branch 1 with a dilation factor of 1 outperforms branches 2 to 7.

**Table 3:** Mean SSIM and mean PSNR (dB) on Set11 with one scale dilated convolution model (after [48])

| Connected branch | All | None | 1 | 2 | 3 | 4 | 5 | 6 | 7 |
|---|---|---|---|---|---|---|---|---|---|
| SSIM | **0.8723** | 0.6756 | 0.8595 | 0.7198 | 0.6973 | 0.7007 | 0.7006 | 0.7000 | 0.6989 |
| PSNR | **29.96** | 25.63 | 29.36 | 26.13 | 25.88 | 25.91 | 25.90 | 25.90 | 25.89 |

Table 4 and Table 5 present additional ablation studies where 2 or 3 branches in Figure 4 are connected while the remaining branches are removed. Branch 1 is retained in all cases due to its outstanding performance, as demonstrated in Table 3. When fusing the features of another branch with branch 1, it is observed that branch 2 has a greater influence compared to the other branches. Furthermore, by incorporating branch 4, the overall performance is further enhanced.

**Table 4:** Mean SSIM and mean PSNR (dB) on Set11 with 2 scales dilated convolution model (after [48])

| Connected branch | 1+2 | 1+3 | 1+4 | 1+5 | 1+6 | 1+7 |
|---|---|---|---|---|---|---|
| SSIM | **0.8650** | 0.8624 | 0.8628 | 0.8625 | 0.8586 | 0.8605 |
| PSNR | **29.57** | 29.51 | 29.51 | 29.50 | 29.32 | 29.41 |

**Table 5:** Mean SSIM and mean PSNR (dB) on Set11 with 3 scales dilated convolution model (after [48])

| Connected branch | 1+2+3 | 1+2+4 | 1+2+5 | 1+2+6 | 1+2+7 |
|---|---|---|---|---|---|
| SSIM | 0.8679 | **0.8683** | 0.8680 | 0.8647 | 0.8664 |
| PSNR | 29.71 | **29.72** | 29.71 | 29.55 | 29.63 |

Table 6 and Table 7 present the performance of the proposed model when some of the 7 branches in Figure 4 are removed. From Table 6, it is evident that branch 1 is the most important, as its removal leads to significant performance degradation. In contrast, the removal of branch 3 and branch 6 has a much smaller impact on model performance. Table 7 provides further insights, demonstrating that branch 2 is more important than branches 4, 5, and 7. Removing branch 2 results in more degradation compared to the removal of the other branches. Overall, these findings highlight the varying importance of the different branches, with branch 1 and branch 2 playing crucial roles in the model's performance.

**Table 6:** Mean SSIM and mean PSNR (dB) on Set11 with 6 scales dilated convolution model (after [48])

| Disconnected branch | None | All | 1 | 2 | 3 | 4 | 5 | 6 | 7 |
|---|---|---|---|---|---|---|---|---|---|
| SSIM | **0.8723** | 0.6756 | 0.6635 | 0.8674 | 0.8720 | 0.8715 | 0.8713 | 0.8724 | 0.8703 |
| PSNR | **29.96** | 25.63 | 25.49 | 29.77 | 29.93 | 29.92 | 29.90 | 29.94 | 29.84 |

**Table 7:** Mean SSIM and mean PSNR(dB) on Set11 with multiscale dilated convolution model (after [48])

| Disconnected | None | All | 3+6 | 3+6+1 | 3+6+2 | 3+6+4 | 3+6+5 | 3+6+7 |
|---|---|---|---|---|---|---|---|---|
| SSIM | **0.8723** | 0.6756 | 0.8718 | 0.6644 | 0.8665 | 0.8703 | 0.8692 | 0.8699 |
| PSNR | **29.96** | 25.63 | 29.91 | 25.50 | 29.69 | 29.83 | 29.76 | 29.80 |

## 5 Conclusion and Future Work

Inspired by the concepts of deep unrolling methods, we present a novel approach called MsDC-DEQ-Net, which combines the deep equilibrium model based on the ISTA algorithm with multi-scale dilated convolution, for image compressive sensing. By mapping a single iteration of the ISTA algorithm to a deep learning block and using it as a deep equilibrium layer, our model maintains a clear interpretability. Extensive experiments demonstrate that the proposed model achieves competitive performance when compared to state-of-the-art CS methods.

In order to leverage the structural diversity originating from the CS domain [13], we incorporate ResNeXt to enhance performance and the SE block to eliminate redundancy and enhance valuable information. Compared to deep unrolling methods, our proposed model significantly reduces the number of learnable parameters by utilizing only one optimization iteration block.

In future research, we plan to explore the robustness of the proposed model and its application in other fields. Additionally, we recognize the importance of high-throughput methods that facilitate information transition by incorporating multiple channels in the input and output of the iteration block.

**Acknowledgements:**

This research was partially funded with a grant through the Natural Sciences and Engineering Research Council (NSERC) of Canada.